\newcommand{\bea} {\begin{eqnarray}}
\newcommand{\eea} {\end{eqnarray}}
\documentstyle[prd,aps,amstex,latexsym,epsfig,amsfonts]{revtex}
\def\edcomment#1{\iffalse\marginpar{\raggedright\sl#1\/}\else\relax\fi}
\reversemarginpar
\begin{document}
\title
{Tacoma Bridge Failure-- a Physical Model}
\author{Daniel Green and  William G. Unruh}

\address{ 
Department of Physics and Astronomy\\
University of British Columbia\\
Vancouver, BC, Canada, V6T 1Z1
\\email: {\tt drgreen@@stanford.edu, unruh@@physics.ubc.ca}}

\maketitle

\begin{abstract}
The cause of the collapse of the Tacoma Narrows Bridge has been a 
topic of much debate and confusion since the day it fell.  Many 
mischaracterizations of the observed phenomena have limited the 
widespread understanding of the problem.  Nevertheless, there has 
always been an abundance of evidence in favour of a negative 
damping model.  Negative damping, or positive feedback, is 
responsible for many large amplitude oscillations observed 
in many applications.  In this paper, we will explain some 
well-known examples of positive feedback.  We will then present 
a feedback model, derived from fundamental physics, capable of 
explaining a number of features observed in the instabilities of 
many bridge decks.  This model is supported by computational, 
experimental and historical data.
\end{abstract}

\section{Introduction}

In teaching a Physics of Music class by one of us, one of the most surprising 
of physical phenomena is the ability of certain physical situations to convert a steady
condition into oscillations. This includes the steady blowing of air through
the reed of a clarinet, through the lips of a trumpet, or over the hole 
of a flute or a beer bottle, or the conversion of the steady pull of a bow across 
a string in a violin into a steady vibration. In some of these cases the
mechanism is clear. Just as with the conversion of a steady voltage supplied 
to an amplifier into the howl caused by feedback of loudspeaker sound to a
microphone, the reed instruments and the violin bow operated under conditions
in which they act as amplifiers-- sources of negative damping in which 
a small input signal is converted into a larger output. 
In the case of the clarinet reed, there is a direct regime of operation 
of the reed where the reed acts directly as an amplifier. If the external
 pressure is high enough (but not so high as to close the reed), 
an increase in internal pressure will open the reed and allow more 
air into the clarinet. A decrease in internal pressure will close the 
reed and allow in less air.
 This negative slope in the pressure-flowrate curve 
acts as a negative damping on any oscillations within the instrument, causing them to increase until the reed enters a highly non-linear regime.
 Similarly, 
in the violin bow, the presence of the rosin on the bow creates a regime of
negative damping, in which the force on the string decreases as the velocity
increases. Again, the effect 
is to amplify any oscillations in the string-- and since the effect is greatest 
on the resonant modes of the string, this tends (if the bow is properly played) 
to create a large amplitude oscillation at the resonant frequency of the string. 

\begin{figure}
\begin{center}
\includegraphics[width=0.9\textwidth]{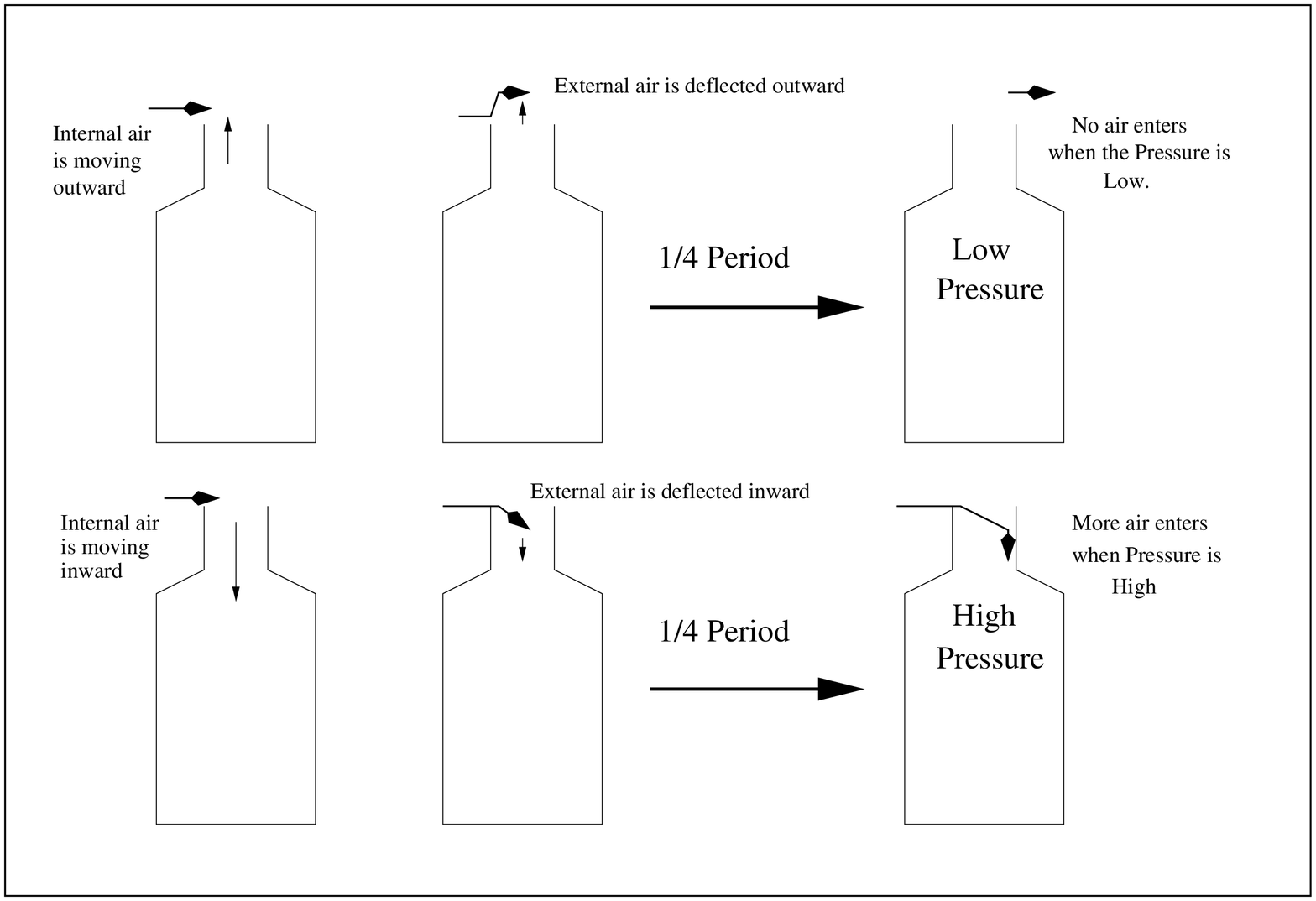}
\caption{The positive feedback that occurs between the internal 
oscillations of a bottle and external air blown over the surface 
at 1/4 the period of the internal oscillation is shown.  
The result is an exponentially increasing oscillation which we hear.}
\label{bottle}
\end{center}
\end{figure}

In the case of the flute, or the coke bottle, the effect is subtler.
There is no direct source for the amplification. Rather it is an interaction
between the oscillation of the air in the instrument, and the air flow 
across the opening (see figure \ref{bottle}). Assume that the time it takes air blown across the 
opening is roughly a quarter the period of the oscillation of the air 
in the instrument.  When the oscillation of the air is out of the bottle, 
the air blown over the surface will be deflected upward, missing the bottle 
at the far edge.  This is correspond with the point in the oscillation where 
the internal pressure is lowest (a 1/4 period later).  Similarly, the deflection 
of the air into the bottle will occur when the pressure is at it's highest.  
The effect of the deflection of the air enhances the pressure in a way that 
is in sync with the pressure itself.  I.e. in this case 
there is a symbiotic relationship 
between the oscillation of the air in the instrument and the feedback mechanism. 
Only those modes with the correct relationship between the oscillation period 
and the time it takes the air to flow across the hole are amplified, have a
negative damping coefficient.

There are two important aspects to note in all of these cases. While the
natural oscillation period of the vibration produced tends to be very close 
to the period of a natural resonance of the instrument (of the air within 
the tube of a clarinet, flute or beer bottle, or of the string of a violin) 
none of these correspond to the traditional notion of resonance. There is no external
oscillating force which is tuned to the natural period of the oscillation.
Rather there is a complex non-linear interaction between the oscillations 
and the external steady action which creates a condition of negative damping (a force proportional to the velocity for the simple harmonic oscillator which is 
the mode of vibration).

One of the most impressive examples of such a conversion of a more or less
steady flow of air into a large amplitude vibration was that of the Tacoma
Narrows Bridge. As Billah and Scanlan complained 13 years ago \cite{BS}, physics text
books insisted in calling this large scale oscillation an example of resonance,
when it was clearly no such thing. Already in the report of the Federal 
Panel \cite{Fed} struck to investigate the collapse, (a report issued only four months after 
the collapse) it was recognised that the collapse was due to another example 
of negative damping, just as in the above musical instruments. However that
report, and much work in the years thereafter have left the physical origin 
of that negative damping unclear. It is clearly due to some aspect of the
turbulence in the flow of the wind across the bridge, but which aspect?
This paper will report of recent work, which attempts to clarify the origin 
of that feedback which resulted in such an impressive example of "music".

The Tacoma Narrows Bridge opened on July 1, 1940 and collapsed on November 
7, 1940 under winds of approximately 40 mph.  During that brief period, it became an 
attraction as it oscillated, at a relatively low amplitude (a few feet)
 in a number of different modes, in all of which the bridge deck remained
 horizontal.  Low mechanical damping allowed the bridge to vibrate for long periods of time. However, 
on November 7th at 10:00 am,  tortional oscillations began that were 
far more violent than those seen previously (measured at one time to have an 
amplitude of about 13 feet, or greater than .7 radians).  At 11:10 am, the midspan 
of the deck broke and fell due to the large stresses induced by the oscillation.  
The H-shape of the bridge cross section was attributed as a crucial component of
the problem. The post-facto study of instability of such bridge sections 
in wind tunnel tests, which formed the core of the Federal Task force report,  resulted in the
realisation that such studies were crucial before building such bridges in the
future.  That the influence of such feedback is still not under control is
evidenced by the very public closing of the Millennium footbridge in London in
2000 one day after opening due to feedback, this time between the bridge and
people walking across it.

At this point, one may ask, "this happened so long ago, bridges today are 
stable, why do we care?" While this may be sufficient for engineers, it is
unsatisfying to a physicist. What exactly is the nature of the feedback which
caused the failure?   In this paper we will examine such a physical model for the 
Tacoma Narrows Bridge.  First, we will revisit 
old explanations that gloss over the underlying problem.  Second, the 
physical model will be presented, followed by evidence in its favour obtained 
from historical, computational and experimental data.

\section{Historical Misconceptions}

 The underlying cause of the collapse of the Tacoma Narrows Bridge has been 
frequently mischaracterized. Although evidence in favour of negative damping 
has been available since 1941 \cite{Fed}, oversimplified and limited theories 
have dominated popular literature and undergraduate physics textbooks. 
Billah and Scanlan \cite{BS} describe and analyse many of the frequently quoted 
explanations.   Most commonly, the collapse is described as a simple case 
of resonance. There exists a video tape (our copy unfortunately or otherwise has
lost its attribution) in which the instability is modeled by a fan blowing on
"wave train", with the demonstrator inserting and removing a large piece of
cardboard from in front of the fan at the resonant frequency of the structure. 
 However, making the comparison to a forced harmonic oscillator 
requires that the wind generate a periodic force tuned to the natural frequency 
of the bridge.  They remark, "texts are vague about what the exciting force 
was and just how ... it acquired the necessary periodicity."  In some cases, 
the periodic vortex shedding of the bridge was used as the source of this 
periodicity (the Van Karman Vortex Street). This assumes that the frequency 
of vortex shedding (Strouhal frequency) matches natural frequency of the 
bridge.  However, the Strouhal frequency of the bridge, under 42 mph wind 
is known to be ~1Hz, far from the observed 0.2 Hz frequency of the bridge 
itself.  The Van Karman Vortex Street could not produce resonant behaviour 
on the day of the collapse.

Although Billah and Scanlan clearly outline the pitfalls of the resonance 
model, recent work \cite{McK} has attempted to resurrect this model using 
non-linear oscillators. For a non-linear oscillator, the frequency is
not independent of amplitude, but can change as the amplitude changes.
This can lead to complex, even chaotic behaviour of the oscillator.

\begin{figure}
\begin{center}
\includegraphics[width=0.9\textwidth]{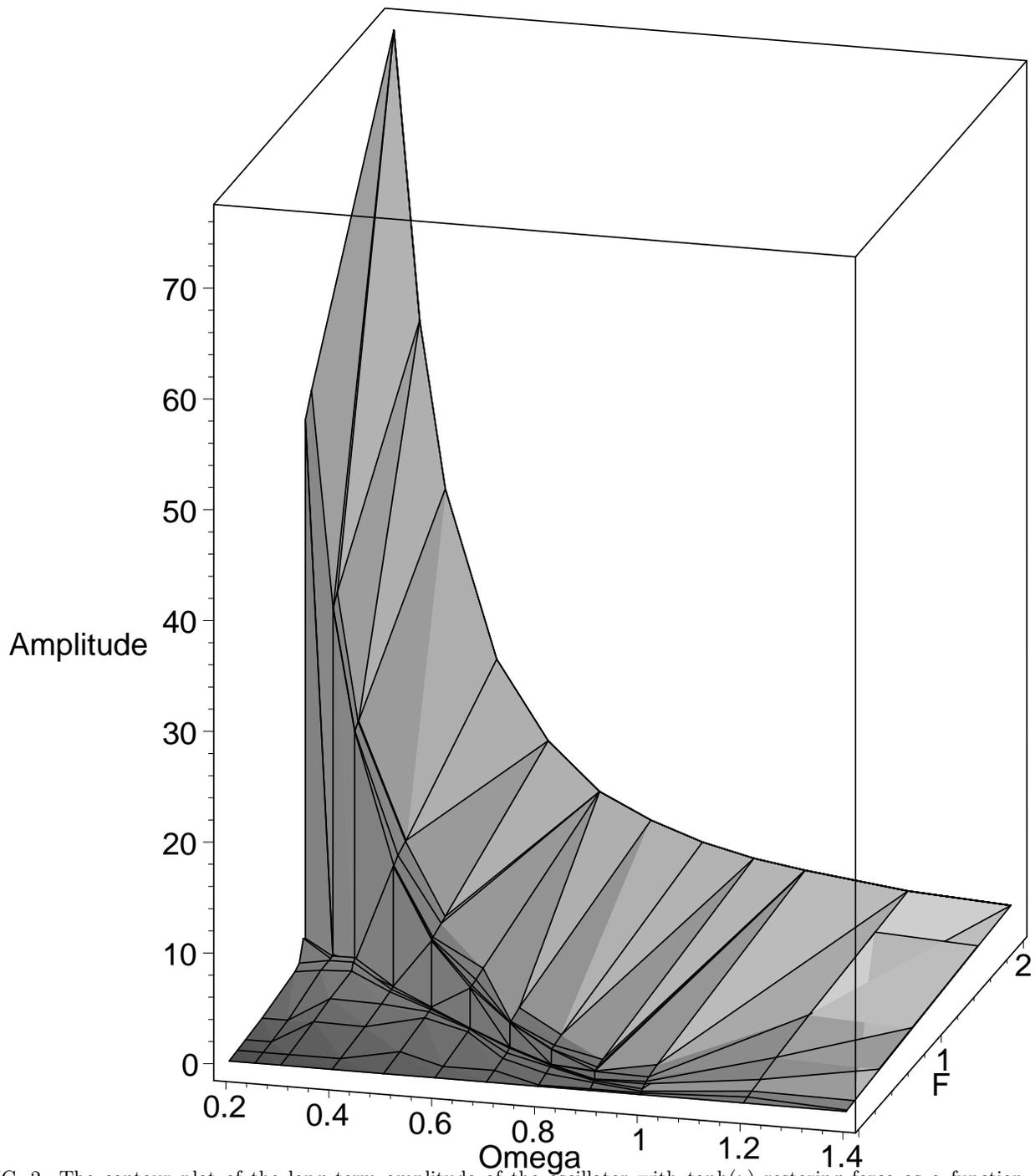}
\caption{
The contour plot of the long term amplitude  of the oscillator with $\tanh(y)
$ restoring force as a function of the frequency and amplitude F of the
driving sinusoidal force.
}
\label{tanh-contour}
\end{center}
\end{figure}

\begin{figure}
\begin{center}
\includegraphics[width=0.9\textwidth]{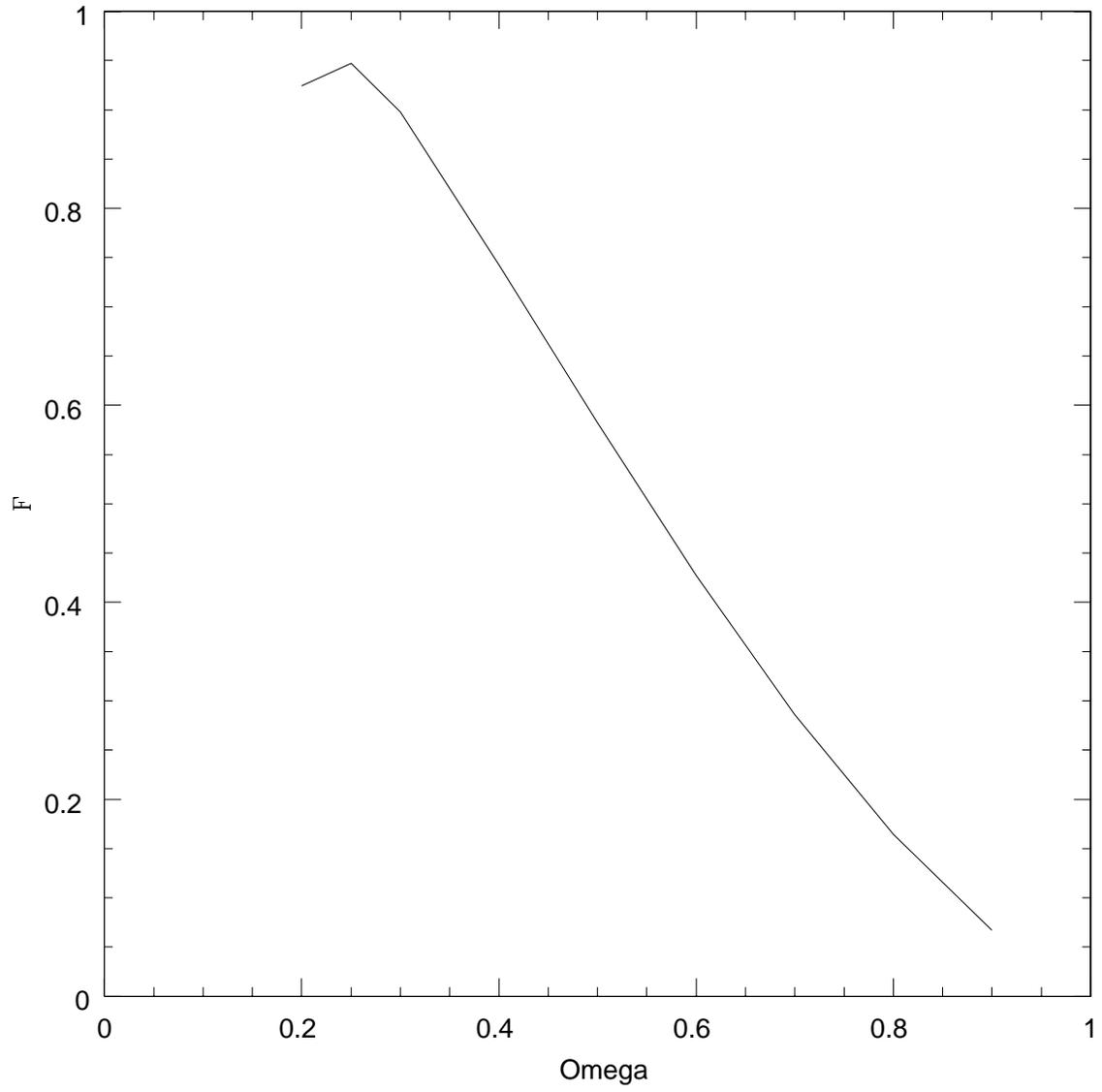}
\caption{
The discontinuity line in $\Omega - F$ plane for figure 2. 
}
\label{tanh-WF}
\end{center}
\end{figure}

\begin{figure}
\begin{center}
\includegraphics[width=0.9\textwidth]{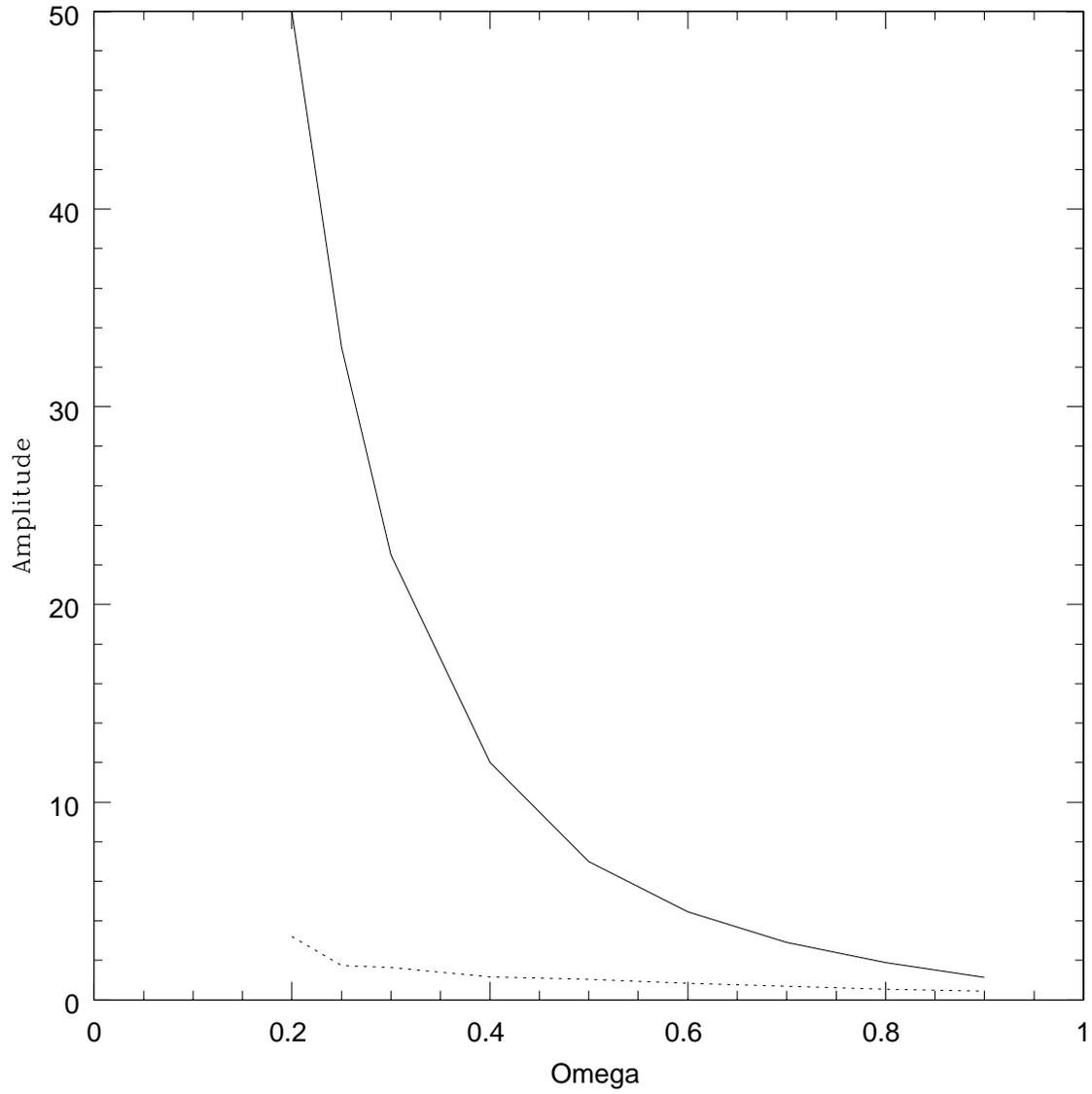}
\caption{
The discontinuity in long term response as a function of $\Omega$ of Figure
2.  The solid line is the value of the long term
amplitude of $y$ across the discontinuity 
at the high value of $F$ while the dotted line is the long term value at the
lower part of the discontinuity.
}
\label{tanh-WA}
\end{center}
\end{figure}

Let us look at the solution for the non-linear equation
\bea
\label{nonlin}
{d^2 y(t)\over dt^2} + .05 {dy\over dt} + \tanh(y(t))=F \sin(\Omega t).
\eea
This is a simple model in which the restoring force goes from the linear
dependence at small amplitudes to a constant restoring force at large
amplitudes.  
In figure \ref{tanh-contour} to \ref{tanh-WA}, we plot some of the features of the long term
behaviour of the  solutions to this equation. In general the system settles
down to a fixed long term approximately harmonic solution (for $\Omega$
very different from the small oscillation value of unity, and F near unity,
one can also get highly non-harmonic chaotic behaviour even at late times,
but we will not go into that here).
There is a discontinuity in the long term solution of this equation as a
function of $F$. Figure \ref{tanh-contour} is the contour plot of the long-term amplitude
as a function of the frequency $\Omega$ and the amplitude of the forcing
term F, while figure \ref{tanh-WF}  is a   plot of $\Omega$ vs $F$ along the line of
discontinuity.  Note that as $\Omega$ goes to 1, the discontinuity disappears. In figure \ref{tanh-WA}, we plot the size of the discontinuity at various values of $\Omega$. The solid line is the value of the long term amplitude of $y$ across the discontinuity 
at the high value of $F$ while the dotted line is the long term value at the
lower part of the discontinuity. In all cases the initial value of $y$ and
$dy\over dt$ is zero. Other integrations showed that the location of the
discontinuity (value of $F$ for given $\Omega$)
can depend on those initial values.

This eliminates one of the problems with the standard physics
explanation of the bridge motion by resonance, namely that the wind
speed would have to be absurdly accurately tuned so that the resonance
condition would produce the incredibly high amplitudes of oscillation.
Once the bridge entered the upper leg of the resonance curve, it could
remain there even as the frequency of the driving force frequency  varied.

This would also explain one puzzling anomaly of the bridge motion on
Nov 7th. Since about 6AM that morning the bridge had been oscillation in
a non-torsional mode, with a relatively low amplitude (1.5 feet or so),
and with about 8 or 9 nodes in the oscillation along the length of the
bridge. Suddenly at around 10AM, while Kenneth Arkin (Chairman,
Washington Toll Bridge Authority) was trying to measure the
amplitude of oscillations with a surveyor's transit,

``...Then the mid span targets disappeared to the right of my vision. 
Looking over the transit, mid span seemed to have blown north approximately 
half the roadway width coming back into position in a spiral motion...''

I.e., the transition to the very large amplitude torsional oscillation was
very abrupt. The bridge basically went from no torsional oscillation to
a very large torsional oscillation almost instantaneously. This seemed
to occur with little change in the wind velocity (the famously quoted
42mph for the wind was measured before the bridge went into its torsional
oscillation from on the deck of the bridge). 

McKenna and coworkers have used this to argue that what happened at that
moment was that the bridge suddenly made a transition from the lower
amplitude solution to the upper, perhaps due to the effect of a gust of
wind. 

  In their model for the bridge, they introduce the non-linearity by,
 instead of using a standard spring 
model for the effect of the cables on the bridge, the effect of the
cables are described by one-sided springs (mimicked by the tanh force function
in (\ref{nonlin})). I.e., when the side of the
bridge rises above its normal equilibrium position, the cables on that
side are assumed to go slack, putting the bridge into free fall
(gravity being the only restoring force). As is well known from
watching a bouncing ball, the frequency of oscillation of an object
under such conditions decreases as the amplitude increases (ie, $\alpha$
in the above is negative).
There is however no indication in any of the reports of the engineers
watching the oscillation. Farquason, one of the chief consultants on
the bridge from the University of Washington, went out onto the bridge
during its violent torsional oscillation, both to observe the behaviour
of the bridge from close up, and to try to rescue the car and dog
abandoned on the bridge. He reports that the riser cables were not
slack during the oscillation \cite{Fed}.

In addition to the original suggestion, McKenna suggests that
perhaps the replacement of the small angle linear approximation for the
motion of the bridge by the 
proper trigonometric functions could provide sufficient non-linearity. 
 Through numerical calculations of the response of such a non-linear
oscillator to a force with constant frequency and amplitude, they argue
 that the non-linearities of the trigonometric functions alone 
can allow for  the same kind of bimodal response, with  large amplitude 
oscillations when the linear case may not. However these would be similar
to the above model near $\Omega=1$ where the jump in amplitude of the
oscillation is not very large.  
 Although these models provide a
more complete analysis of motion, their  analysis ignores the cause 
of the driving forces entirely.  In particular, it is assumed that there exist
  sinusoidal forces on the bridge 
with constant period and amplitude chosen, purely for their ability to drive the
resonance and  without regard for their possible 
physical origin.  However, even if such a bimodal response is a
component in understanding the detailed
of the response of the bridge,  without understanding the origin of these forces, 
little insight into the cause of the collapse is gained by introducing 
the non-linearities.

In general, the above models do not address the wealth of data collected 
in wind tunnels and on the day of collapse.  In particular, the logarithmic 
decrement of the oscillation (effective negative damping coefficient)
 had already  been measured over an extensive range of  
wind speed in wind tunnel tests immediately after the collapse, and is
reported in the Federal Task Force report.  These results reveal that
the wind induced damping coefficient changes from positive  
 to negative at a critical wind speed.  Furthermore, above this critical
wind speed, the behaviour of the negative damping coefficient 
appears to be linear in  the wind velocity.  One should be capable of explaining 
these and other results with a realistic model.  Most importantly, one should be 
able to describe what is different in the case of the Tacoma Narrows Bridge 
that caused it to collapse.  Since non-linearities are present in
 all suspension bridges, why do   other bridges remain stable even in
higher winds?  
Although non-linear models may provide some insight into the motion of large span 
bridges (we in particular remain unconvinced of their importance),
 they do not address these physical concerns.   

\section{Vortices Again?}

A key clue to the bridge collapse is given in a few frames of the films
which were taken by E. Elliot of the Camera Shop in Tacoma\cite{film}. At one point
the concrete in the bridge deck begins to break up and throws dust into the
air. This dust acts as a tracer for the airflow over the bridge. In
figure \ref{mvfrm} we can just see a large vortex
moving across the bridge beyond and to the right of the car abandoned on
the bridge (the development in the movie is much clearer). The vortex is
first observed at just about the point where the bridge has levelled out in
its oscillation, and by the time the bridge reaches its maximum
counter clockwise excursion, the vortex has fallen apart and moved off the
right edge of the bridge.

Vortices form in the wake of an oscillating body from two different sources.  
First of all, the Von Karman Vortex Street forms at a frequency determined 
by the geometry and the wind velocity.  These vortices form independently 
of the motion and are not responsible for the catastrophic oscillations 
of the Tacoma Narrows Bridge.  Vortices are also produced as a result of 
the body's motion.  In most cases, the frequency of vortex formation matches 
the frequency of the oscillation. The behaviour of these vortices will 
depend on the geometry and the motion of the body.  The nature of these 
motion-induced vortices played a central role in the collapse of the bridge.

Kubo et al. \cite{Kubo} were the first to observe the detailed structure of the 
wake that forms around the bridge's H-shaped cross-section (H-section) 
and for a rectangular cross section (which is also prone to violent tortional 
oscillations).  In both cases a regular pattern of vortices appeared on 
both sides of the bridge deck.  They speculated that the spacing between 
consecutive vortices was the likely cause of different vertical and tortional 
modes of oscillation observed during the brief lifetime of the Tacoma Narrows 
Bridge.

\begin{figure}
\begin{center}
\includegraphics[width=0.9\textwidth]{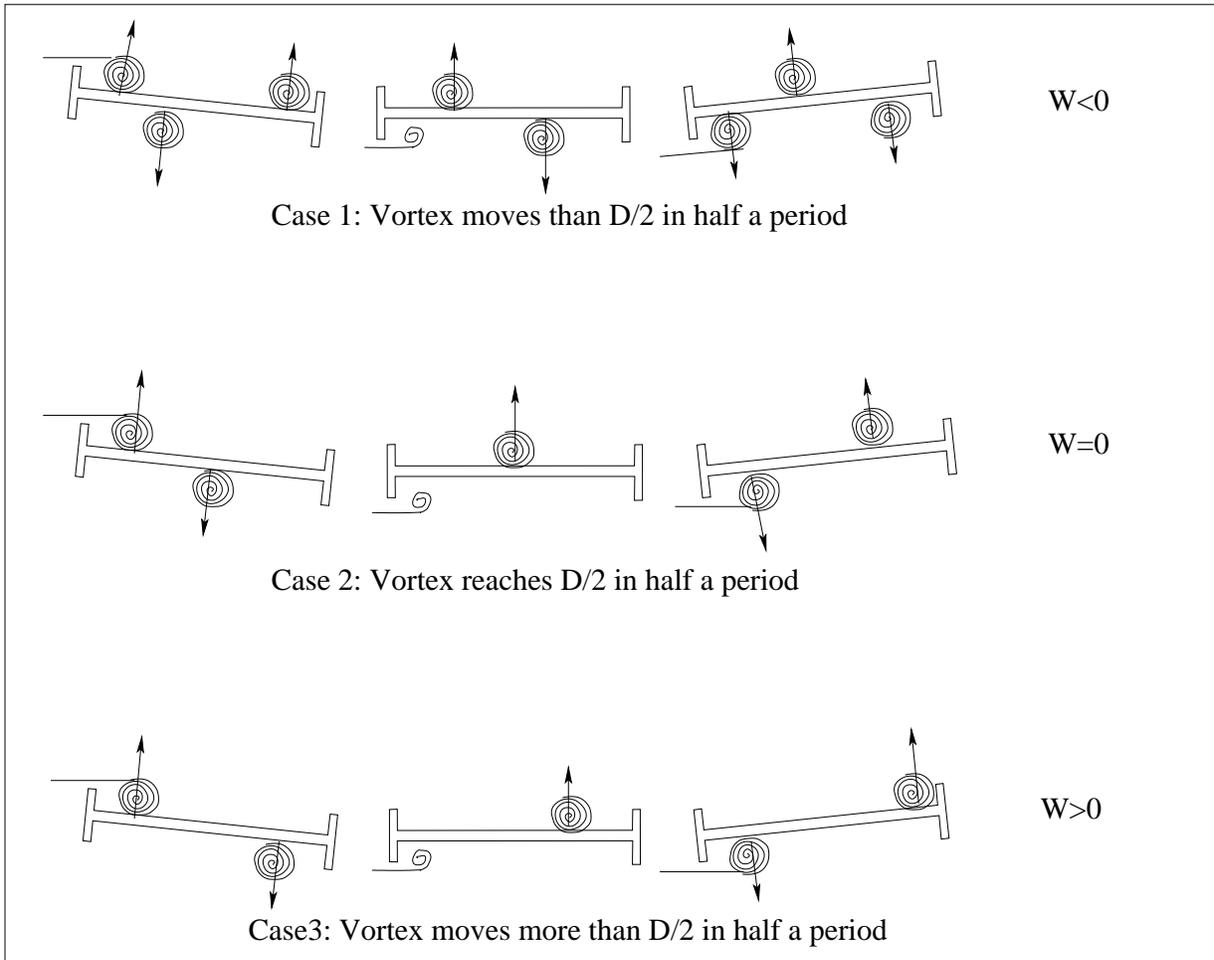}
\caption{Larsen's model and analysis of the feedback mechanism.  The 
work can be calculated from the force time the velocity of the deck 
at the point of the force.  When the vortex drifts across the 
deck in exactly one period, the force it produces does no work 
on the bridge.  This is the critical wind speed.  Above this wind 
speed the work is positive.  At slightly slower wind speeds, 
this work is negative.  Using a drift speed of ${U\over 4}$, 
${U_{c} P\over D}$ is 4.}
\label{Larsen}
\end{center}
\end{figure}

Based on the observations of Kubo et al, and his own computer simulations, 
Larsen \cite{Lars} produced the first physical model of the bridge collapse.  
Since this is a model of positive feedback, let us assume that the bridge 
is oscillating at its resonant frequency with some amplitude.  In this 
model, a vortex is formed at the leading edge of the deck, on the side 
in the direction of motion, as the angle crosses zero (see figure \ref{Larsen}).  I.e., the vortex
forms suddenly at the front edge of the bridge just as the bridge passes
the horizontal position, on the shadowed side of the deck.  Since 
a vortex is a low-pressure region, a force in the direction of the vortex 
is produced.  Each time the bridge is level, another is vortex is generated.  
Once the vortex forms, it will drift down the deck of the bridge producing 
a time-dependant torque.  Here Larsen makes two assumptions based on his 
observations at dimensionless wind speeds near $UP/D=4$, where $U$ is
the wind velocity, $P$ is the period of oscillation, and D is the width
of the bridge.  First, the vortex 
drifts at a constant speed of roughly $0.25U$.  Secondly, he assumes that 
the force produced by the vortex is independent of time.  

There are a number of attractive features of the Larsen model.  Primarily, 
it can provide some insight into the problem with only a few simple assumptions.  
Larsen analyzed this model by considering the work generated by vortices 
as they drift over the bridge.  Figure \ref{Larsen} shows the three cases he considered.  
The first, at winds less than the critical wind speed, the vortices do 
not cross the entire bridge in one-period and produce forces that will 
dampen the oscillation. At the critical wind speed, the vortex crosses 
the bridge in exactly one period.  The resulting work is zero.  At higher 
wind speeds, the vortex crosses the entire bridge in less than a period, 
doing work on the bridge.  As a result, the bridge would gain energy and 
the amplitude of oscillation would grow.  If the drift speed is between 
${1\over 4}U$ and ${1\over 3.6}U$, then the critical wind speed ${U_cP\over
D}=3.6-4.0$, consistent 
with measurements.  Kubo et al. showed that this same pattern could be observed 
at lower wind speeds and may explain the other observed modes of oscillation.  
This is an important result.  Using only a simple model, the critical wind 
speed has been recovered.  Further simulations by Larsen show that by preventing 
these vortices from forming, by replacing the solid trusses with perforated 
ones, the critical wind speed is increased by several times.

\begin{figure}
\begin{center}
\includegraphics[width=0.9\textwidth]{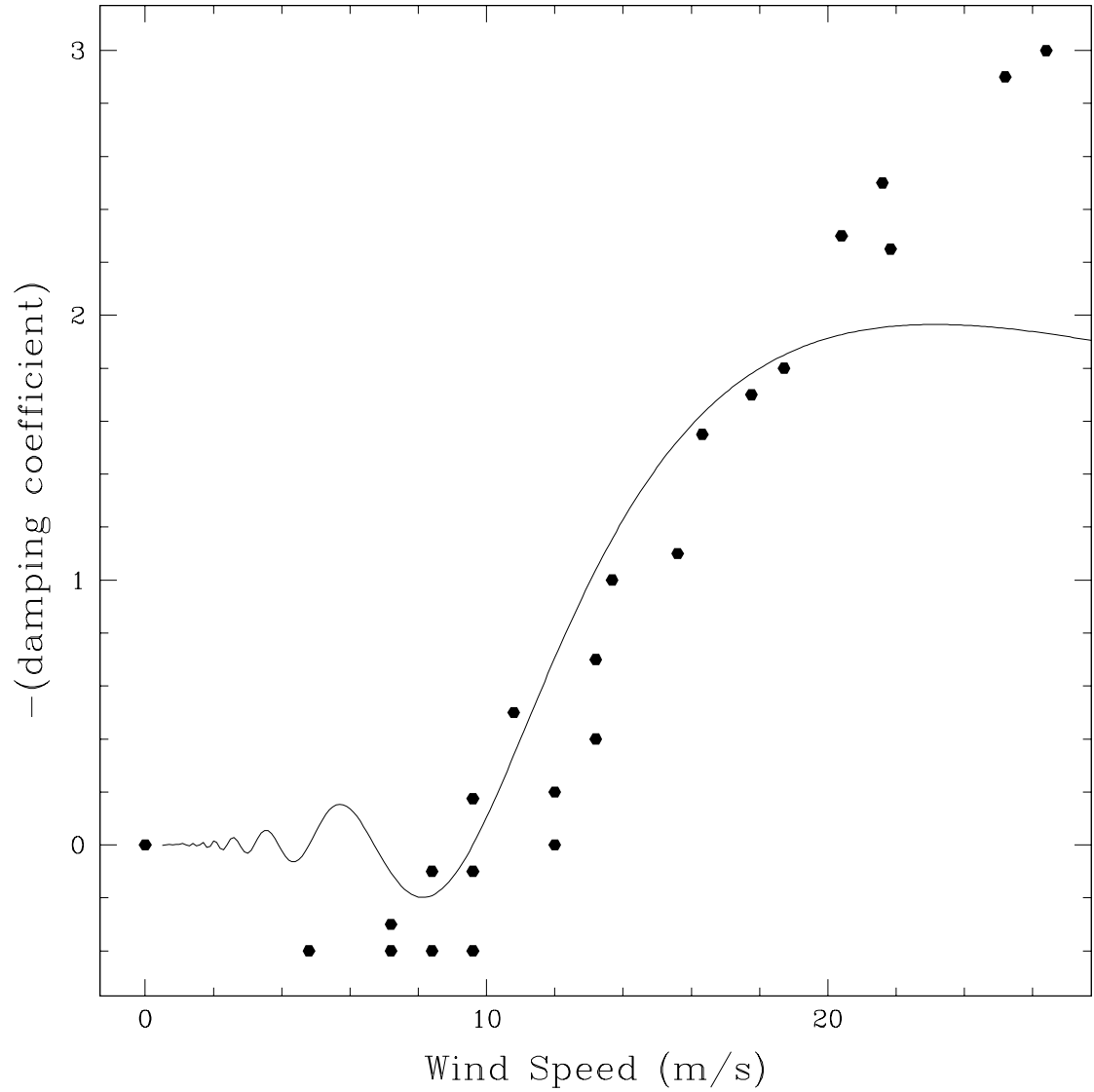}
\caption{Additional analysis of Larsen's model is 
compared to wind tunnel data.  Although the 
model agrees with the data around the critical wind speed, problems 
occur at higher wind speeds.}
\label{Larsen2}
\end{center}
\end{figure}

Despite this success, this analysis is somewhat incomplete given the data 
available.  In particular, it does not address the wind speed dependence 
of the damping as a function of wind speed.  Let us consider this in the 
context of the Larsen model.  Working in dimensionless units, 
 the drift time of the vortex across the deck is
$k=3.6-4.0$.  
Therefore we can write, 
\bea
\alpha =A \sin(2\pi t/P) \\
\dot\alpha = {{A2\pi} \over P}\cos({{2\pi t} \over P}),
\eea
where A 
is the amplitude.  We can also write the torque as
\bea
 T=F(1-2t/k)=F(1-t/2) 
\eea
where $F$ is the force generated by the vortex.  Taking the inner product 
of $\dot\alpha$ and the torque (this is just the work, $\int_{t=0}^t F\cdot v dt$, written in a suggestive form), 
\bea
<T|\dot\alpha>=\int_0^k(T\dot\alpha)dt=-F[\sin[{2\pi\over
U_r}]+{U_r\over\pi}(\cos[{2\pi\over U_r}]-1)],
\eea
where $U_r={U\over U_c}$.  Based on Bernoulli's equation, one might assume that $F$ is proportional to $U^2$.  The resulting behaviour of the energy
feed in by a vortex to the bridge as a function of $U_r$ is shown in 
figure \ref{Larsen2}.  This clearly does not fit the behaviour of the bridge at high wind 
speeds.

The combined work of Larsen and Kubo et al. suggests that this view may 
be accurate (or close enough) at low wind speeds.  However, a number of 
questions remain after analysing this model.  First of all, how do the 
vortices form and how to they drift?  If they do move at 1/4 U, why?  Secondly, 
is the force constant as a function of time?  Finally, can we adjust this model to reproduce the wind speed dependence of the damping?

\section{Potential Flow Models}

We will examine several potential flows in order to address the previous 
questions.  First of all, we will look at how vortices drift near boundaries.  
Secondly, we will consider how a vortex drifts near the trailing edge of 
the bridge.  Finally, we will examine the production of vortices at the 
leading edge.

\begin{figure}
\begin{center}
\includegraphics[width=0.9\textwidth]{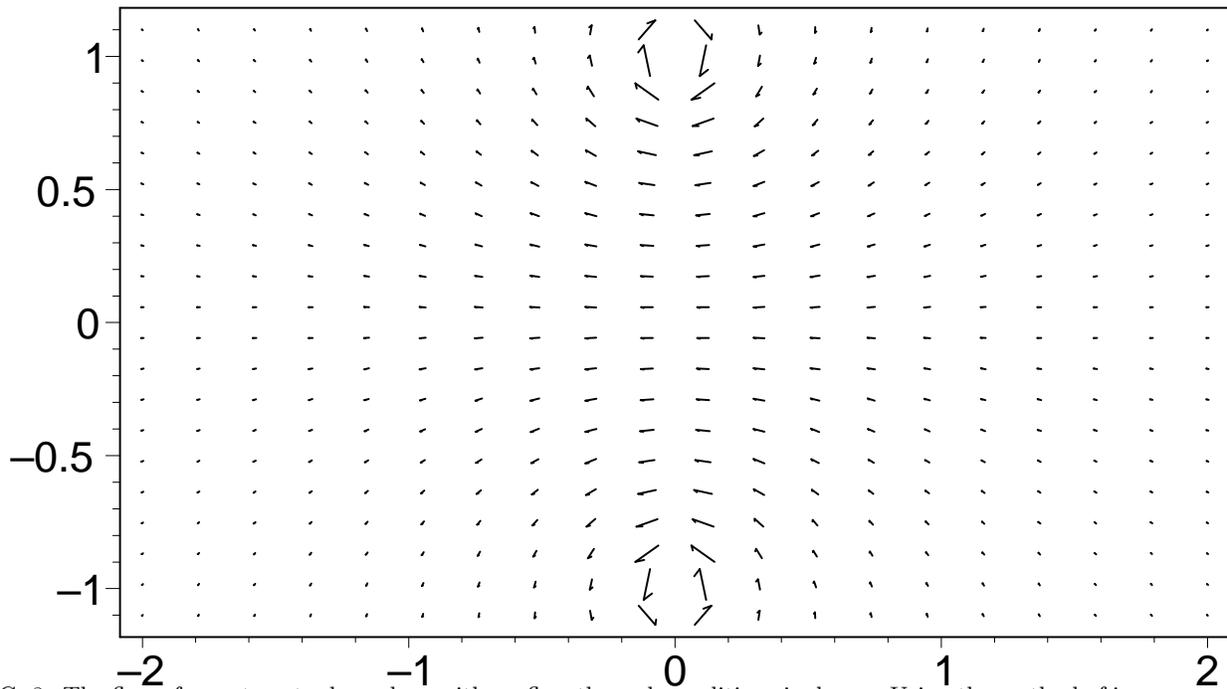}
\caption{The flow of a vortex at a boundary with no-flow 
through conditions is shown.  Using the method of images, a 
counter vortex is placed across the boundary (y=0).  Notice 
that the fields in the y direction cancel along the line y=0.}
\label{Image}
\end{center}
\end{figure}

Potential flows are solutions to the vector Laplace equation.  As such, 
uniqueness of solutions allows one to use the method images to solve simple 
problems.  For example, a vortex at a solid boundary (no flow 
through conditions) can be solved using a vortex of the opposite orientation 
reflected across the boundary (see figure \ref{Image}).  This same model will apply when 
the vortex is also placed in a uniform flow.  The vorticity transport equation 
in two-dimensional incompressible flow will have the form 
\bea
{{\partial \omega}\over {\partial t}}+ (u \cdot \nabla) \omega = \nu {\nabla}^2 \omega,
\eea
where $\omega=\nabla\times{u}$ is the vorticity and $\nu$ is the viscosity.
Since $\nu$ is small, the vortex will drift at the speed of the 
fluid at its centre (ignoring it's own contribution to the flow).  Therefore, in this case 
the vortex will drift at 
\bea
v_d=U-{\gamma\over (4\pi a)}
\eea
 where $\gamma $ is the vortex 
strength and $a$ is the distance from the boundary to the centre of the vortex.  
Therefore, a vortex on the bridge deck will drift at a reduced speed under 
an external flow.

When the vortex reaches the back edge of the bridge, the local fluid may 
not be along the bridge deck, as assumed above.  Any laminar-like flow 
over the back edge must separate from the bridge in order to get over the 
back truss.  This will cause the vortex to move off the bridge deck.  This 
will reduce the force generated by the vortex at the surface.  As a result, 
the force at the back edge of the deck should be lower than elsewhere on 
the bridge. Also, as the vortex and the separation line leave the back edge,
the reduced pressure within the vortex will pull in fluid from the lower side
of the deck, often producing a counter flow vortex beneath the detaching
original vortex.

\begin{figure}
\begin{center}
\includegraphics[width=0.9\textwidth]{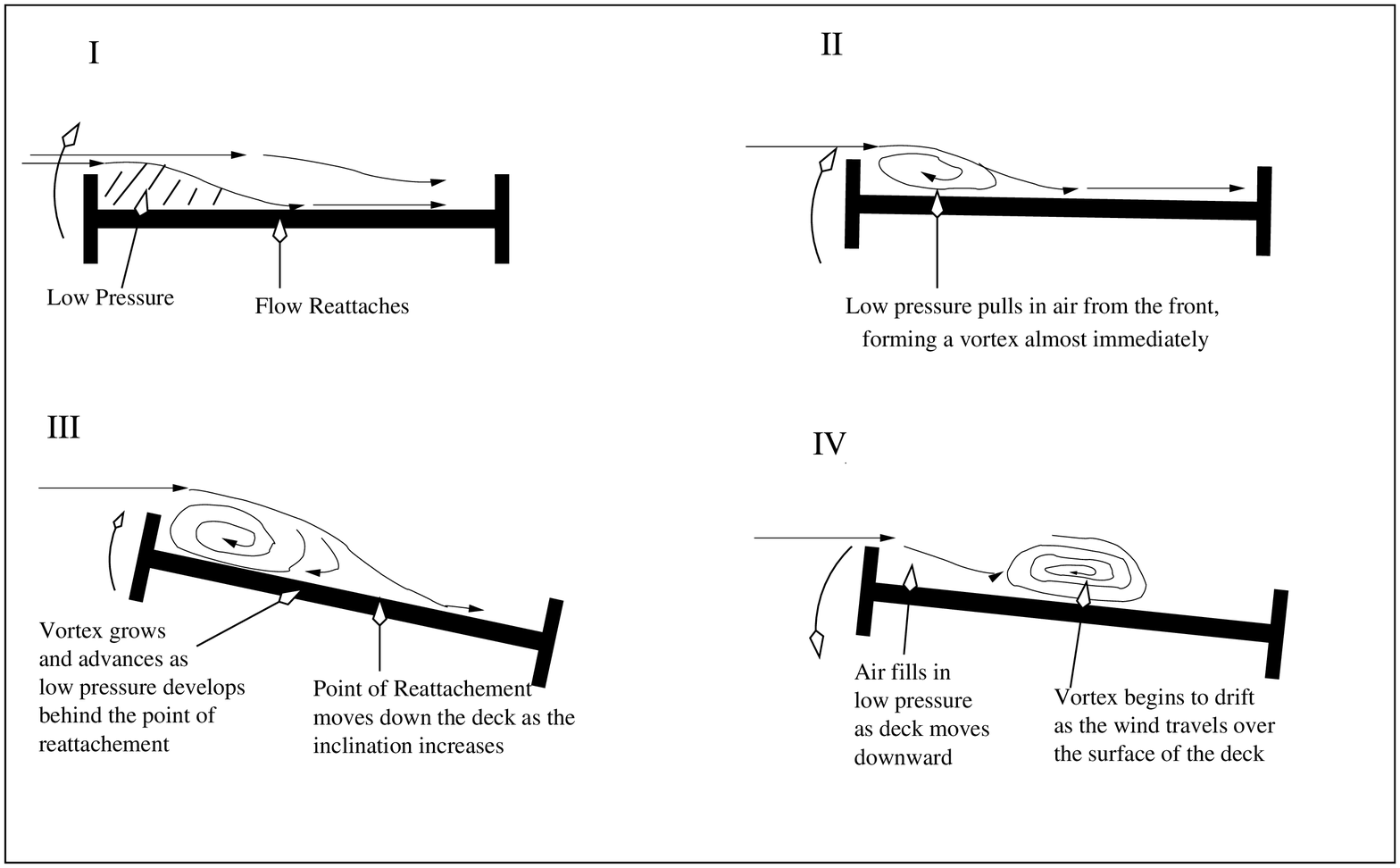}
\caption{Vortices are formed in the low-pressure wake of 
the front edge of the bridge deck.  As the bridge deck rises, 
the point where the flow reattaches moves down the deck.  The 
low pressure region that forms behind this point causes the 
vortex to grow and move.  When the deck begins to move downward, 
air begins to fill in this low pressure region.  This wind will 
also cause the vortex to drift off the deck at a speed related 
to the angle of the bridge.}
\label{model}
\end{center}
\end{figure}

Although the formation of the vortices may seem intuitively obvious, it 
provides insight into the large wind speed problems seen in the Larsen 
model.  First let us consider what happens in the static case at the onset 
of an external flow.  The fluid will first flow around the solid truss 
at the leading edge, separating from the boundary (see figure \ref{model}).  Due to the reduced density,
a low-pressure region will form behind the truss that will 
force the gas downward (the $\nabla \cdot U=0$ condition will have the same effect 
in the incompressible case).  Eventually, gas will be pulled backwards 
into this region producing a vortex.  When the bridge is not moving, the 
vortices remained fixed behind the truss.  Furthermore, equal strength 
vortices form on either side of the deck.  Therefore, the drifting of the 
vortices is a result of the motion of the deck.  As the deck moves upward, the size 
of the low-pressure region increases on the top of the deck.  During the 
upward part of the motion, the point on the deck where the flow reattaches 
will move progressively down the deck, producing a large low-pressure region 
from the truss to this point.  Once the deck begins moving back down, the 
vortex will find itself in a region that looks progressively more laminar.  
This will push the vorticity off the back edge.  The speed at which this 
occurs will depend on the angle.  However, it will increase the pressure 
on the deck and separate the vortex from the front edge.  For large wind 
speeds, all vortices will be pushed off the back edge by 3/4 of a period 
as the entire deck is exposed to strong laminar flow.

The opposite side of the bridge is affected in the opposite way.  As the 
deck moves, the point of reattachment moves back towards the truss.  This 
fills in the region where the vortex would otherwise form.  At best a very 
small vortex will form.  As a result, we have a large vortex that forms 
on one side and a small vortex that forms on the opposite side.

After considering these different aspects of the vortex formation and motion, 
we get somewhat different results than the simple linear assumptions made 
by Larsen.  First of all, the motion across the bridge deck will not be 
uniform.  In the first quarter of the period, the vortex grows do to the 
movement of the point of reattachment.  Since the vortex is always touching 
the front truss, this means that the torque will have the same sign as the 
angular velocity for this first quarter period.  The exact behaviour of 
the torque will depend on the wind speed.  When the bridge reverses it's 
direction of motion, the vortex will begin to drift down the bridge.  The 
local wind velocity will govern the drift speed.  This will depend on the 
angle of the bridge and the strength /position of the vortex.  Finally, 
the force generated by with vortex will drop as in moves off the back 
edge.  The net result is that the torque is large at the beginning of the 
period when the angular velocity is high.  In the second part of the motion, 
the magnitude of the force decreases, giving less of an impact when it 
is out of phase with the angular velocity of the bridge.

At very high wind velocities, the size of the vortex grows to the size 
of the bridge deck.  At this range the behaviour can once again be simplified.  
As the bridge moves, it creates a low pressure region which grows to cover the 
entire deck.  The resulting pressure looks like a step function 
that moves across the bridge until a uniform low pressure is established.  
When the wake moves beyond the back edge, flow entering from the opposite 
side  of the bridge to fill in the low pressure region, dissipating the 
pressure almost uniformly.  As a result, the pressure will decay without 
producing many large asymmetries along the length of the bridge.  
As a result, the torque will remain approximately zero.

In order to generate a single model on all scales, capable of describing 
the data, we combined the above formation process into the sum of a point 
vortex and a step function.  Linear and quadratic wind speed dependence 
was used for the force generated by the vortex and step 
function respectively.  Such a model is based on the separation of 
the low pressure that generates the vortex and the reduced pressure generated 
by the vortex itself.  This will allow us to recover both models on appropriate 
scales.  The constant pressure distribution cause by the separation moves 
along the deck at the speed of the front edge of the vortex (U/2) while 
the vortex itself is centred on the region and thus moves at U/4.  The 
particular model is motivated by the two factors.  First of all, it describes 
the additional torque due to the extended low-pressure region behind the 
truss.  As a result, the damping coefficient is calculated using
\begin{eqnarray}
<T|\dot\alpha>&=&A[u^2 \int_0^{2D\over u}((D^2-(t u/2-D)^2)\dot\alpha)dt+10u\int_0^{4D\over u}((D-t u/4)\dot\alpha)] \\
\nonumber
&=& -{5\over 8\pi^2} [u^3 (24\pi\cos({48\pi\over 5u} )-
5u\sin({48\pi\over 5u} )+24\pi) + 2\pi u(48\pi\sin({96\pi\over 5u} )+
5\cos({96\pi\over 5u} )- 5u)]
\end{eqnarray}
where D is the width of the deck.  This precise model is also motivated by our observations of the 
pressure in numerical calculations described in section V.  In particular, the relative value of 10 between the two terms was chosen by observation of the relative influence of the vortex and step-function at different wind speeds.

The results with this model are shown in figure \ref{qmodel}.  The asymmetry in the 
torque generated by the growing phase allows us to recover the asymptotic 
linear behaviour that is expected.  Furthermore, the critical wind speed 
is found to be consistent with experimental values.  By adjusting the overall 
amplitude, the model is found to be consistent with data found in \cite{Fed}.

It should be observed that the linear model Larsen uses is a low wind speed 
approximation to this model.  If the wind speed is suitably low, the vortex 
remains on the bridge for at least one full period.  As a result, the period 
where the vortex is growing is less significant given the long period of 
interaction.  Furthermore, the differences in drift speed become less significant 
because the variation is occurring on a period much shorter than the period 
that the vortex is on the bridge.  Finally, the points of reattachment 
and separation are much closer to the trusses at low speeds, reducing the 
effects seen at both ends.  The end result is that a time averaged force 
and speed would give essentially the same results under these conditions.  
However at speeds in excess of ${UP\over D}=5-6$ this approximation will likely 
not hold.

\section{Historical and Computational evidence}

We previously presented a model based mainly on simplified laminar flows.  
However, more rigorous evidence is required to support this model rather 
than just the fit to the wind speed data.  It must be shown that such a 
flow appears when one calculates the full flow numerically or during experiments.  
In particular, we want to show the high wind velocity flow is different 
from that observed by both Larsen and Kubo et al.

\begin{figure}
\begin{center}
\includegraphics[width=0.9\textwidth]{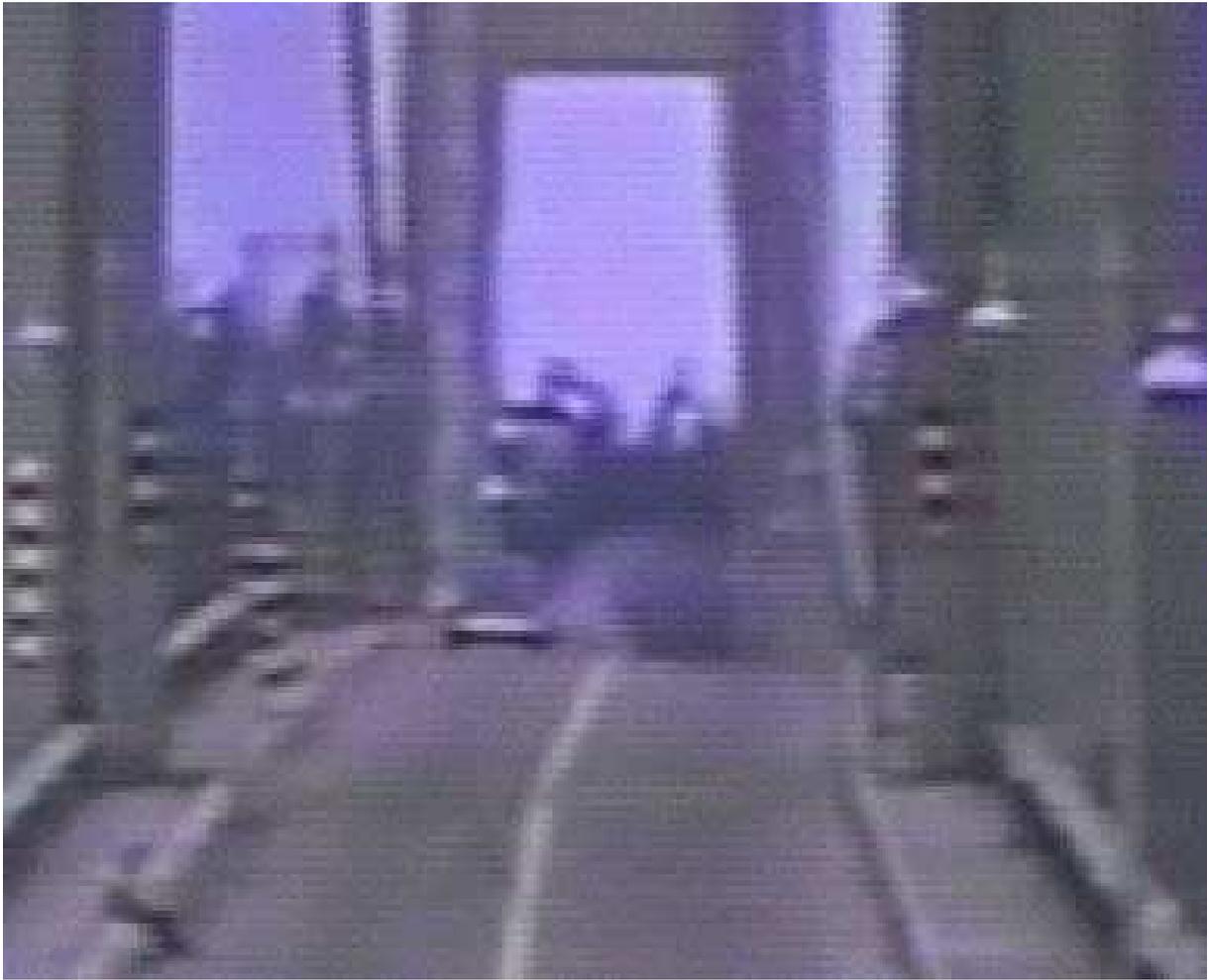}
\caption{
Frame from the film taken by Elliot et al of the bridge collapse. To the right of the car
is a large vortex outlined by the cement dust from a section of the roadway
that was apparently disintegrating. To the left sits Prof. Farquarson. The
behaviour of the vortex is much clearer in the film itself. The roadway is
almost level in its counter clockwise rotation. Copyright "The
Camera Shop". Used with permission.
}
\label{mvfrm}
\end{center}
\end{figure}

The most exciting, and useful evidence comes from the film coverage of 
the November 7, 1940 incident.  At one point during the oscillation, the 
roadway appears to break up throwing  cement dust into  
 the air.  In several frames (figure \ref{mvfrm}) from the original movie, one such incident is seen.  The cement and dust become caught in a vortex that drifts along the bridge 
deck at a rate similar to that seen in simulations (it is not possible 
to determine the exact rate relative to the wind velocity since the instantaneous 
wind velocity in unknown).  This video evidence supports two important 
points in this argument: (1) the existence of drifting vortices on the 
deck, and (2) the vortex crosses the midway point  in less than one period, a requirement 
for negative damping.

Unfortunately, the engineers present at the scene did not risk life and 
limb to get a more complete set of wind, pressure or turbulence measurements.   
As a result, that one observation captured on film seems to be the only
 historical evidence from the 
bridge itself.  Therefore, we are forced to rely on numerical simulations 
to fill in the gaps.  Our simulations were conducted using VXF FLOW produced 
by Guido Morgenthal \cite{GM}.  The code uses discrete vortex methods to 
determine the flow.  The advantage of vortex models for this problem (compared 
to finite element methods) is that it solves for the vortex transport
explicitly. Finite element methods tend to introduce an effective viscosity due
to the finite grid mesh. Since the Reynolds number of the flow across the bridge
was so high (of order $10^6$) the high grid viscosity  could  seriously change
the physics of the modeled flow. Finite element methods offer the advantage of
removing such grid viscosity, modeling the flow by a finite distribution of
individual vortices. While such methods suffer from the relatively poor
convergence properties of finite element methods (tending to converge at a rate
which goes as the $\sqrt{N}$ where $N$ is the number of vortices), the ability of
modern computers to handle large numbers (our simulations typically had more
than $10^5$ vortices in play at any one time), and its ability to model the high
Reynolds numbers typical of the bridge without the introduction of artificial
viscosity, made using such discrete vortex methods  the preferred procedure. 
The generosity of Morgenthal in allowing us to use his code in this
investigation was thus critical to the any success which we had.

Our simulations were conducted by manually oscillating the deck at some 
frequency and amplitude and observing the resulting fluid flow.  These 
tests covered a large range of incident wind speeds and amplitudes of oscillation.  
At each time step, the location and velocity of every vortex was returned 
along with the pressure along the surface of the bridge.  From the pressures, 
the lift, drag and torque are calculated.  Since the vortex drifts at the local 
wind speed, the vortices can be used to visualize the fluid flow.

At wind speeds below 10 m/s, the pressure and fluid flow are very 
similar to those described by Larsen.  In particular, a vortex is generate 
at the front and drifts along the deck.  This vortex seems to make the 
largest contribution to the pressure.  When the vortex is created a more 
extended low pressure region is created behind the truss as the bridge rises.  
As the vortex begins to drift, the extended constant pressure slowly disappears 
leaving only the vortex.

\begin{figure}
\begin{center}
\includegraphics[width=0.9\textwidth]{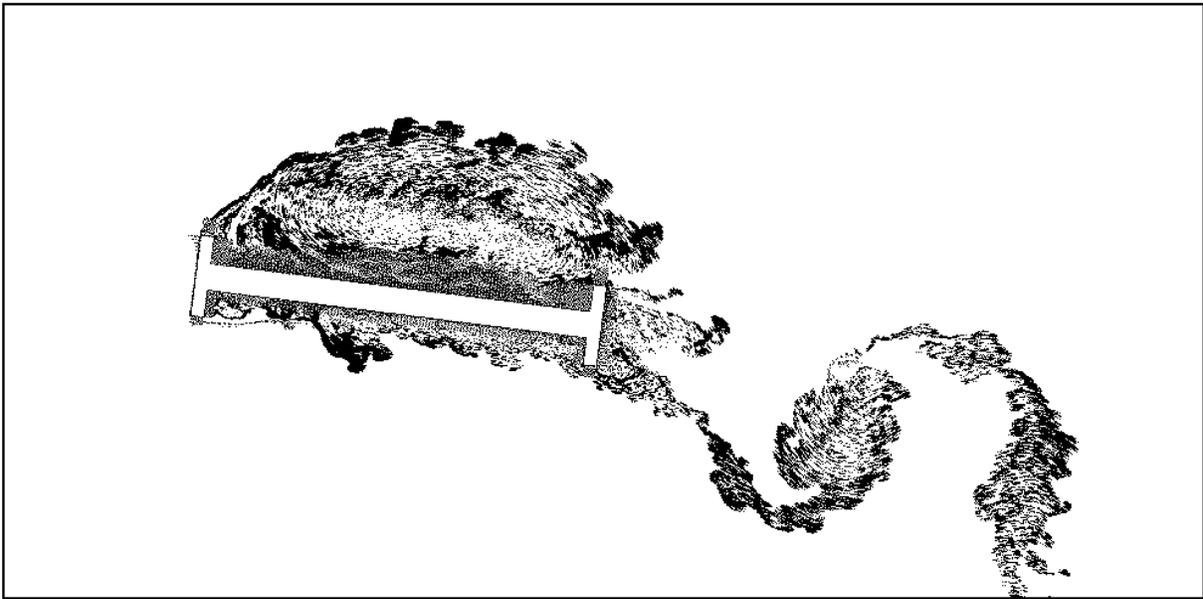}
\caption{Frame from the simulation using VXF Flow at 19 m/s.  There is clearly a large vortex over halfway across the bridge as in becomes level.  This is very similar to the frame from the original movie.  There is also a low-pressure region that covers the length of the deck, which is consistent with the model presented in this paper.}
\label{frm19}
\end{center}
\end{figure}

At wind speeds around the 19 m/s (42 mph) that were observed on the day, 
turbulence begins to play a more significant role in the pressure.  As the 
deck rises the extended pressure covers over half of the bridge deck.  A 
large vortex forms near the front of the region.  The vortex makes an equal 
contribution to the extended constant low pressure region behind the truss.  
As the vortex drifts off the deck, a low pressure covers the deck.  
This motion varies with the angle of the bridge and does not appear 
constant.  The flow is turbulent as smaller vortices are shed behind 
the first.  The pressure remains low until the deck is exposed entirely 
to the wind, at 3/4 of a period when the angle is maximum in the 
opposite direction.  The similarity between the simulation and the original film can be seen in figure \ref{frm19} where a frame form the simulation is shown at the same point as the video frame (figure \ref{mvfrm}).

At wind speeds above 30 m/s, turbulence dominates the flow.  As the deck rises, 
the low pressure moves across the deck, covering it entirely before the deck 
reverses direction.  A single vortex is formed at the front of this region a 
similar to the slower wind speeds but it's contribution to the pressure isn't 
noticeable.  However, unlike the slower wind speeds, this is not the only 
large vortex that forms.  Vortices are formed off the front edge at a high 
frequency that drift a varying speeds.  At times, these vortices can catch 
up with the front vortex.  Large counter vortices are formed at the back 
edge as air is pulled from the far side of the bridge to fill in the low 
pressure.  All these large vortices interact through magnus forces and produce 
unpredictable fluid behaviour.  Such a highly turbulent flow introduces rapid 
noticeable changes in the pressure.  This makes torque and work calculations 
highly erratic.  However, because the vortices are shed at a high frequency 
relative to that of the bridge, the contribution of the noise to the work 
over many periods will be negligible.

\begin{figure}
\begin{center}
\includegraphics[width=0.9\textwidth]{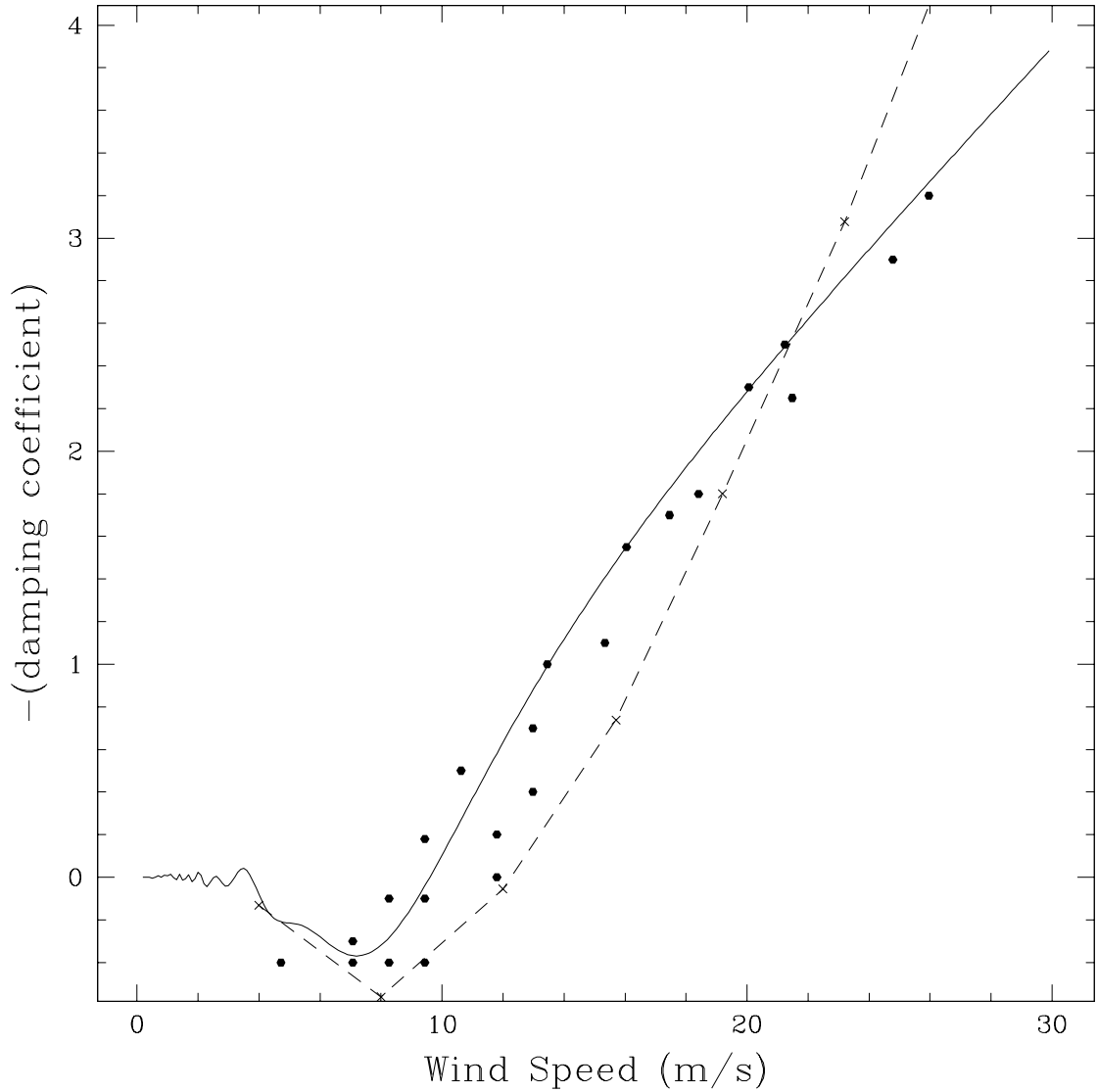}
\caption{
The results of our model are compared to data 
taken from wind tunnel tests (points) and simulations using VXF Flow (crosses and dotted line).  This model does not 
suffer from the large wind speed deviations seen in 
Larsen's model.  It also successfully describes low wind 
speed behaviour as well.  The simulation, which guided the model, is
further from the experimental data.  The error on the simulated data
is likely substantial, since we were never able to simulate more than
10 periods.  The simulated damping coefficients shown are the average over all periods after the first.}
\label{qmodel}
\end{center}
\end{figure}

The torque calculations were used to generate effective damping coefficients 
over a range of wind speeds (figure \ref{qmodel}).  Over a range of wind speeds from 4 to 30 m/s, 
the damping coefficient at .3 radians are similar to those 
measured in the Federal Report.  At wind speeds above 30 m/s, the negative damping coefficient begins to drop.  Particularly at high wind speeds, the 
work is very sensitive to the time stepping and the time integration.  For 
example, by decreasing the time step we can be many points onto or near the 
line seen at lower wind speeds.  Mogenthal noticed this problem as well.  
His simulations of other bridge section had a turn over in the damping 
inconsistent with measurements.  Therefore, it is unknown whether the drop 
in the negative damping is a physical or computation effect.  The differences between the experimental and simulated data in figure \ref{qmodel} are likely due to error.  Because of the noise in the data and the few periods we were able to simulate, error could be as high as 30 to 50 percent for some points.  Furthermore, the experimental data was taken when the bridge is allowed to oscillate out of control, unlike the simulation.

\begin{figure}
\begin{center}
\includegraphics[width=0.9\textwidth]{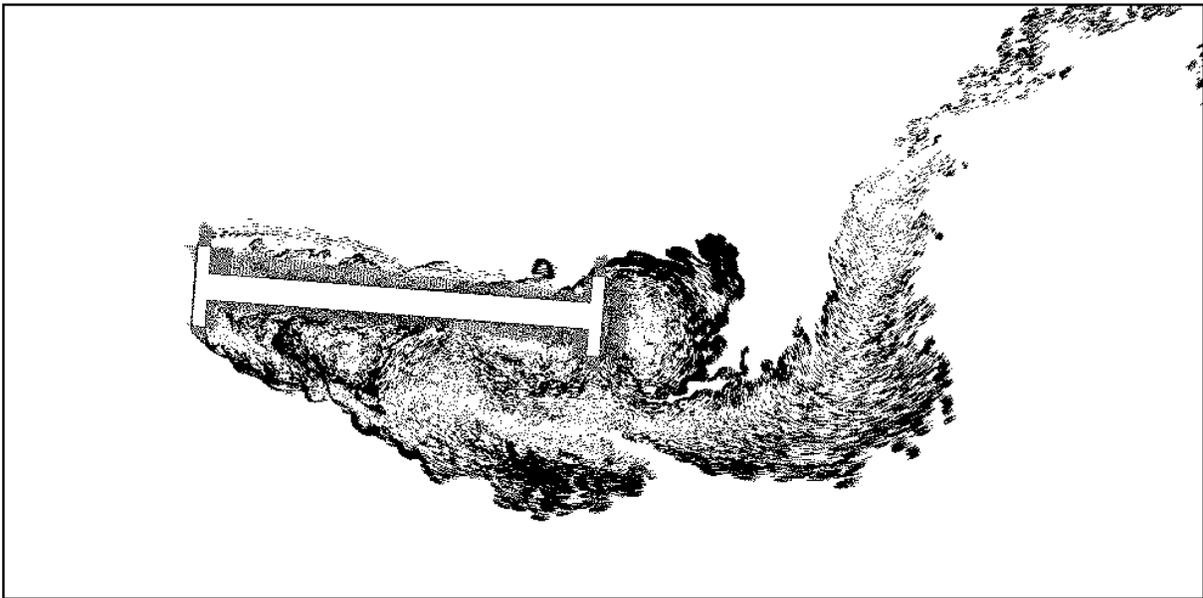}
\caption{The flow over the bridge cross section as the deck rises, as calculated with VXF Flow, at a wind of 23 m/s.  Each point represents a point vortex.  (Top) A vortex begins to form behind the front edge as a low-pressure region forms.  This low pressure is due to the separation of the moving fluid from the boundary of the deck.  As the deck rises this region will get larger. (Bottom)  The vortex that formed on the bottom is being blown off the deck.  The pressure on the bottom is dropping uniformly.}
\label{frame1}
\end{center}
\end{figure}

\begin{figure}
\begin{center}
\includegraphics[width=0.9\textwidth]{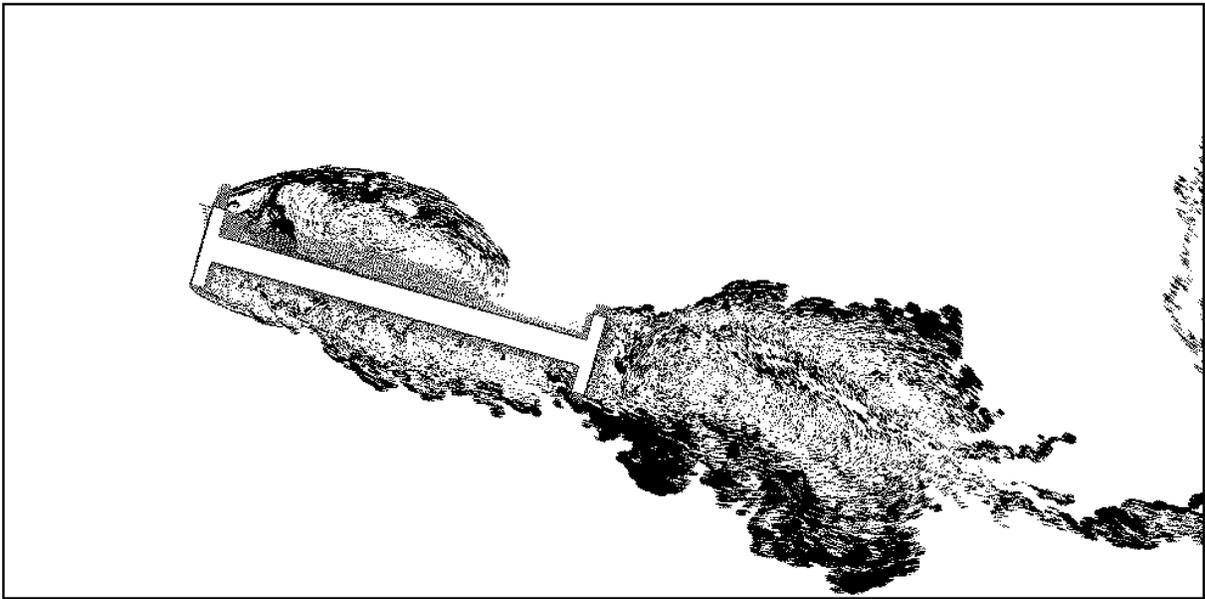}
\caption{The flow over the bridge cross section halfway through the rise of the front edge, as calculated with VXF Flow, at a wind of 23 m/s.  Each point represents a point vortex.  (Top)  The vortex has grown to cover half of the deck.  Because the deck is still rising, the low-pressure region extends from the front of the deck to backmost edge of the vortex.  (Bottom) The bottom of the deck has now been completely exposed to laminar flow.  Few vortices remain.
}
\label{frame2}
\end{center}
\end{figure}

\begin{figure}
\begin{center}
\includegraphics[width=0.9\textwidth]{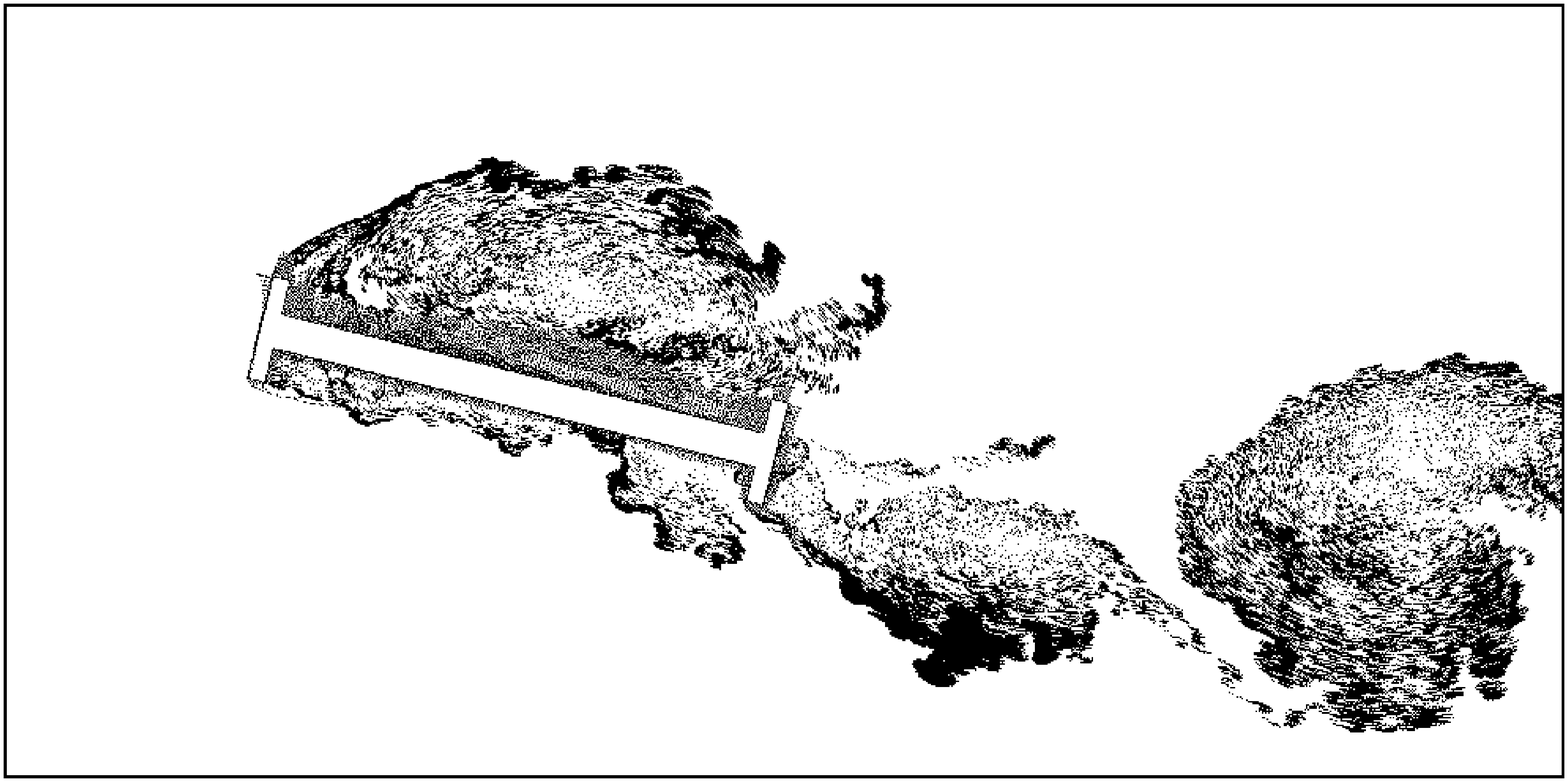}
\caption{ The flow over the bridge cross section after the full rise of the front edge, as calculated with VXF Flow, at a wind of 23 m/s.  (Top) As the deck descends from the maximum height, the vortex is now being to be pushed off the back edge.  The vortex has grown to cover a large part of the bridge.  (Bottom) Laminar flow dominates since the bottom has been completely exposed to the wind. }
\label{frame3}
\end{center}
\end{figure}

Given these observations, how does our model compare?  The generic description 
of the physics is consistent with the behaviour of the bridge over the 
period of the bridge (ignoring the high frequency noise).  The vortex 
forms in a large low pressure region and separates from it as the bridge 
moves downward.  A simplification of this occurs at very high wind speeds when the low pressure covers the entire bridge before the motion reverses.  
Furthermore, the pressure generated by the vortex seems to decay somewhat at the 
far edge of the bridge.  The advantage of this model can be seen in figures \ref{frame1} to \ref{frame3}.  The Larsen model does not adequately explain data or simulations at around 23 m/s.  However, the fluid flow in these figures is in reasonable agreement with our model.  

However, the specific mathematical model is not entirely 
consistent with the general behaviour.  The difficulty is that there is no obvious 
mathematical description of the torque generated by this behaviour 
that is consistent with a large range of wind measurements.  The mathematical 
model was taken to be consistent with the constant pressure motion at 
high wind speeds and a single vortex at low wind speeds.  The 
difficulty is that the behaviour of the associated with this model in 
the intermediate wind speeds many not be accurate.  In particular, at 
some intermediate wind speeds (around 10 to 14 m/s), the constant part 
of the pressure (associated with the moving front of the wake) does not 
cover the entire bridge.  However, the model uses a step function that 
crosses the bridge nonetheless.  Furthermore, motions are still taken to 
be at constant velocities.  This is clearly not the case, but will 
average out to something near constant.

The reason that further modifications to this model have not 
been made is that any improvement will require a number of 
new parameters (there are already four parameters in this model).  
Furthermore, assumptions about the manner in
 which the pressure drops and the vortices move must be made.  
With such modifications, little is change is observed in the 
damping coefficient.  Additional problems with making these 
changes are associated with making a universal wind speed dependent 
model.  Although a modification may be made according to 
observations at one wind speed, the wind speed dependence 
of the model is not restricted.

To this point, the amplitude dependence of the force has been 
assumed to be linear (a requirement for the to be proportional to $\dot\alpha$).  
This was tested using torque measurements over a range of 
amplitudes at a fixed wind speed.  The work done in a period generally 
increases with amplitude.  Over a range of amplitudes from 0.1 to 0.25 radians this 
linear assumptions appears to be valid.  At lower wind speeds this is not a 
certainty.  The difficulty in determining the behaviour is due to the large 
amount of noise generated by other features of the flow.  This problem is most 
significant at low amplitudes where the torque can be overshadowed 
by other effects. Although these effects will average out to zero, 
the limited time of a simulations doesn't ensure that these contributions 
are zero.  As a result, the error in measurement of work becomes too large to 
constrain the low amplitude dependence.

\section{Conclusion}

Einstein once said "Everything should be made as simple as possible- but 
no simpler".  For every problem, one looks to explain a wide class of observed 
phenomena with a single basic cause.  In trying to understand fluid structure 
interactions, like the Tacoma Narrows Bridge, accomplishing this goal is 
non-trivial.  Fluid mechanics has typically been the domain of 
experimentalists since the 
governing equations are notoriously difficult to solve.  As a result, extracting 
any underlying simplicity is done without rigorous proof.  Consequently, 
many attempts to explain instabilities have been rejected on the basis 
that they are too simple to be the true cause.  Nevertheless, these unsuccessful 
attempts should not discourage researchers into believing that complex 
unpredictable forces that are generated in high Reynolds flows are at the 
heart of the matter.

The model presented is another step along the path towards the goal of 
a comprehensive single model of the phenomena observed the morning of November 
7, 1940.  The majority of physical data can be adequately explained using 
the vortex-induced model.  The detailed method through which the oscillatory 
behaviour is establish may require some further details, but it seems that 
the main features of the wind structure interaction over a large range 
of wind speeds can be explained in the context of our vortex formation 
model.  However, the range of wind speeds where the model is applicable 
has not been fully established.  At the extreme high and low values, computational 
calculations become less reliable and experiments are difficult.  At very 
low wind speeds vortices may cease to form, while at high wind speeds the 
structure of the wake may loss all periodicity.  Nevertheless, this model 
seems appropriate over a large enough range that it should be useful for 
many  applications, and in particular may give a physics lecturers a model to
replace the naive and wrong "resonance" model so often used in undergraduate
lectures.

\begin{thebibliography}{99}
\bibitem {Fed} O.H Ammann, T. Von Karman and G.B. Woodruff.
 ``The failure of the Tacoma Narrows Bridge'' Report to the 
Federal Works Agency, 28 March 1941
\bibitem {BS} K.Y. Billah, R.H. Scanlan. ``Resonance, 
Tacoma Narrows bridge 
failure and undergraduate physics textbooks'' Am. J. Phys. 
59 (2), 118-124 (1991)
\bibitem {McK} P.J. McKenna. ``Large Torsional Oscillations in 
Suspension Bridges Revisited: Fixing an Old Approximation'' 
Am. Math. M., January 1999
\bibitem{film} Barnet Elliott, Harbine Monroe, Aug. von Boecklin
(1940),``The Collapse of the Tacoma Narrows Bridge" Film available in
VHS, PAL, and DVD from The Camera Shop, Tacoma Wa. Frames from the film are
used with permission of the copyright holder.
\bibitem {Kubo} Y. Kubo, K. Hirata, K. Mikawa. ``Mechanism of 
aerodynamic vibration of shallow bridge girder section'' J. 
Inus. Aero. Wind Eng., 42, 1297-1308 (1992)
\bibitem {Lars} A. Larsen. ``Aerodynamics of the Tacoma Narrows 
Bridge - 60 Years Later'' Struc. Eng. Intern. 4, 243-248 (2000)
\bibitem {GM} G. Morgenthal. ``Aerodynamic Analysis of Structures 
Using High-resolution Vortex Particle Methods'' Ph.D thesis, 
Cambridge University, October 2002.
\end{thebibliography}
\end{document}